\documentstyle[prl,eqsecnum,aps]{revtex}

\begin{document}
\author{J. Q. Shen\footnote{E-mail address: jqshen@coer.zju.edu.cn}}
\address{Centre for Optical
and Electromagnetic Research, State Key Laboratory of Modern
Optical Instrumentation; Zhejiang Institute of Modern Physics and
Department of Physics, Zhejiang University, Hangzhou 310027, P. R.
China}
\date{\today }
\title{Monomode photon spin operators projected onto the fixed frame and quantum-vacuum geometric phases of photons inside a noncoplanar optical fibre\footnote{The present paper is a supplement to the two published papers\cite{Shen1,Shen2}. It will be submitted nowhere else for the publication, just uploaded at the e-print archives.}}
\maketitle

\begin{abstract}
The propagation of monomode photons inside a coiled optical fibre
was regarded as a time-dependent quantum evolution process, which
gives rise to a geometric phase. It is well known that the
investigation of non-adiabatic geometric phases ought to be
performed only in the Schr\"{o}dinger picture. So, the projections
of photon spin operators onto the fixed frame of reference is
discussed in this paper. In addition, we also treat the
non-normal-order spin operators and consider the potential effects
({\it e.g.}, {\it quantum-vacuum geometric phases}) of quantum
fluctuation fields arising in a curved optical fibre. The
quantum-vacuum geometric phase, which is of physical interest, can
be deducted by using the operator normal product, and the doubt of
validity and universality for the normal-normal procedure applied
to time-dependent quantum systems is thus proposed. In the
Appendix, the discussion of possible experimental realizations of
quantum-vacuum geometric phases is briefly presented.


\end{abstract}
\section{Introduction}
It is well known that geometric phases appear only in systems with
time-dependent Hamiltonians (or in systems whose Hamiltonians
possess some evolution parameters). We can calculate the adiabatic
geometric phase (Berry's topological phase) of time-dependent
systems by means of Berry's phase formula, an exquisite expression
which opened up new opportunities for investigating the global and
topological properties of quantum evolutions\cite{Berry2}. In
order to calculate the non-cyclic non-adiabatic geometric phase
one can, generally speaking, exactly solve the time-dependent
Schr\"{o}dinger equation by making use of the Lewis-Riesenfeld
invariant theory\cite{Lewis}.

In this paper two subjects on the wave propagation of light in a
noncoplanarly ({\it e.g.}, helically)
 curved optical fibre are discussed: the projection of photon spin in a comoving frame of reference
 onto the fixed frame and the geometric phases of photons at quantum-vacuum level. Although the {\it quantum-vacuum geometric phases} in the
 fibre, which arises from the
 zero-point electromagnetic fields of vacuum, has been calculated in one of our papers\cite{Shen2}, we think this problem still deserves further
 detailed discussions in the present paper. In an attempt to find an experimental realization of
 quantum-vacuum geometric phases ({\it i.e.}, extracting quantum-vacuum geometric phases from the
 total geometric phases of left- and right- handed light), we meet, however, with difficulties to
 achieve this aim, since due to the opposite signs in helicity eigenvalue of left- and right-
 handed circularly polarized lights, the quantal phases at vacuum level is eliminated automatically
 by each other and consequently we cannot easily detect them in experiments. In the Appendix, we continue to consider
 the problem of how to test the {\it quantum-vacuum geometric phases} of light in the noncoplanarly curved optical fibre.
 We hope much attention would be attracted to this subject both
 theoretically and experimentally.

Investigation of quantum-vacuum geometric phases due to vacuum
fluctuation energies possesses the theoretical significance: in
quantum field theory the infinite zero-point energy of vacuum is
often cancelled (deleted) by using the normal-product procedure
and we thus re-define the vacuum backgrounds. It is believed that
this formalism is often valid for the {\it time-independent}
quantum systems since in these cases the divergent background
energies may have no observable effects and hence do not influence
the physical results of the interacting quantum fields involved.
This, however, may be no longer valid for the {\it time-dependent}
quantum systems (such as the quantum fields in the time-dependent
gravitational background, {\it e.g.}, the expanding universe),
since in these cases the time-dependent zero-point fields of
vacuum will also participate in the time evolution process and
therefore cannot be regarded merely as an inactive onlooker ({\it
i.e.}, a simple passive background). So, it is necessary to
consider the validity problem of normal-order procedure in {\it
time-dependent} quantum field theory. We think, in the literature,
this problem gets less attention than it deserves.
\section{Monomode photon spin operators projected onto the fixed frame of reference}
It is essential to think of the {\it non-adiabatic non-cyclic}
geometric phases in the Schr\"{o}dinger picture. If one deals with
geometric phase problem in, {\it e.g.}, the Heisenberg picture,
then the wavefunction will possess a time-dependent phase factor
which cannot be determined by equation of motion itself. So, when
considering the geometric
phases\cite{Shen1,Shen2,Chiao,Tomita,Gao} of photons inside a
noncoplanarly curved optical fibre, it is of physical significance
to study the projections of photon spin operators onto the fixed
frame of reference. In what follows we will discuss this quantum
mechanical problem.

The spin angular momentum operator of photon fields reads (in the
unit $\hbar=1$ )
\begin{equation}
S_{ij}=-\int{(\dot{A}_{i}A_{j}-\dot{A}_{j}A_{i})}{\rm d}^{3}{\bf
x}, \label{eq21}
\end{equation}
where one can expand the three-dimensional electromagnetic vector
potentials ${\bf A}({\bf x},t)$ as a Fourier series\cite{Bjorken}
\begin{equation}
{\bf A}({\bf x},t)=\int{{\rm d}^{3}{\bf
k}\frac{1}{\sqrt{2(2\pi)^{3}\omega_{\bf k
}}}\sum_{\lambda=1}^{2}{\vec
\varepsilon}(k,\lambda)[a(k,\lambda)\exp(-i{k\cdot
x})+a^{\dagger}(k,\lambda)\exp(i{k\cdot x})]},    \label{eq a}
\end{equation}
where the frequencies $\omega_{\bf k }=c|{\bf k }|$, and
${\vec\varepsilon}(k,1)$ and ${\vec \varepsilon}(k,2)$ are the two
mutually perpendicular real unit polarization vectors, which are
also orthogonal to the wave vector ${\bf k}$ of electromagnetic
wave. Note that here summations with respect to $\lambda$ are over
polarization states $\lambda=1, 2$ (for each ${\bf k}$).

Insertion of the expression (\ref{eq a}) for ${\bf A}({\bf x},t)$
into (\ref{eq21}) yields
\begin{equation}
S_{ij}=-i\int{{\rm d}^{3}{\bf
k}[\varepsilon_{i}(k,1)\varepsilon_{j}(k,2)-\varepsilon_{j}(k,1)\varepsilon_{i}(k,2)][a^{\dagger}(k,1)a(k,2)-a^{\dagger}(k,2)a(k,1)]}
\label{eq22}
\end{equation}
with $a^{\dagger}(k,\lambda)$ and $a(k,\lambda)$ being the
creation and annihilation operators of polarized photons,
respectively, in the comoving frame of reference. Note that here
the photon spin operators in (\ref{eq21}) is of a normal-order
form, {\it i.e.}, $S_{ij}=-{\bf
:}\int{(\dot{A}_{i}A_{j}-\dot{A}_{j}A_{i})}{\rm d}^{3}{\bf x} {\bf
:}$, where the creation and annihilation operators appearing in
{\it normal order} are so arranged: the latter is placed to the
right of the former, which will theoretically cancel (or delete)
the vacuum quantized fields. Thus the operator product under the
normal order sign (denoted by double-dot symbols $:$ $:$) can also
be called {\it normal product}. To discuss the quantized
electromagnetic fields at quantum-vacuum level, we should consider
the non-normal order of spin operators. In Sec. III, we will
analyze the quantum-vacuum fluctuation contributions to geometric
phases of photon fields propagating inside a curved fibre. It is
shown that both left- and right- handed (LRH) circularly polarized
light in the noncoplanar fibre will possess a so-called {\it
quantum-vacuum geometric phase} (but the signs of these two vacuum
phases are just opposite, namely, if the vacuum phase of
right-handed circularly polarized light is positive, then that of
left-handed polarized light has a minus sign\footnote{Thus the
quantum-vacuum geometric phases of left- and right- handed
polarized light are often cancelled with each other, and it is
therefore not easy for physicists to measure them in experiments.
This problem, which I have considered for three years (2000-2003),
will be discussed in the Appendix.}). This fact will be discussed
in more detail in Sec. III, where we will give a firm theoretical
background for {\it quantum-vacuum geometric phases}.

Since for the planar wave, the following mathematical requirement
is satisfied ({\it i.e.}, the transverse nature of electromagnetic
wave)
\begin{equation}
{\vec\varepsilon}(k,1)\times {\vec \varepsilon}(k,2)=\frac{{\bf
k}}{k},
\end{equation}
the expression (\ref{eq22}) for photon spin operators may be
rewritten
\begin{equation}
{\bf S}=-i\int{{\rm d}^{3}{\bf k}\frac{{\bf
k}}{k}[a^{\dagger}(k,1)a(k,2)-a^{\dagger}(k,2)a(k,1)]}.
\end{equation}
Here the creation and annihilation operators
$a_{j}^{\dagger}(k',\lambda')$ and $a_{i}(k,\lambda)$ of polarized
photons agree with the commuting relation
$[a_{i}(k,\lambda),a_{j}^{\dagger}(k',\lambda')]=\delta^{3}({\bf
k}-{\bf
k}')\delta_{ij}\varepsilon_{i}(k,\lambda)\varepsilon_{j}(k',\lambda')$.
Note that for the case of discrete {\bf k}, the magnetic vector
potentials may be rewritten\cite{Lurie}
\begin{equation}
{\bf A}({\bf x},t)=\frac{1}{\sqrt{V}}\sum_{\bf
k}\frac{1}{\sqrt{2\omega_{\bf k}}}\sum_{\lambda}{\vec
\varepsilon}(k,\lambda)[a(k,\lambda)\exp(-i{k\cdot
x})+a^{\dagger}(k,\lambda)\exp(i{k\cdot x})],    \label{eeqA}
\end{equation}
where $V$ denotes the volume of a cubic enclosure in which
electromagnetic fields are present. In this case, photon spin
operator (\ref{eq22}) is rewritten as ${\bf
S}=-i\frac{V}{(2\pi)^{3}}\int{{\rm d}^{3}{\bf k}\frac{{\bf
k}}{k}[a^{\dagger}(k,1)a(k,2)-a^{\dagger}(k,2)a(k,1)]}$ (here
${\bf k }$ tends to be continuous) with the commuting relation
$[a_{i}(k,\lambda),a_{j}^{\dagger}(k',\lambda')]=\delta^{3}_{{\bf
k},{\bf
k}'}\delta_{ij}\varepsilon_{i}(k,\lambda)\varepsilon_{j}(k',\lambda')$
being satisfied. Thus the monomode photon spin operator (with
discrete ${\bf k}$) in the comoving coordinate system is given as
follows

\begin{equation}
 {\bf S}=-i\frac{{\bf
 k}}{k}[a^{\dagger}(k,1)a(k,2)-a^{\dagger}(k,2)a(k,1)].
\end{equation}
Generally speaking, in an infinitely large space, the summations
in (\ref{eeqA}) with respect to ${\bf k}$ are over all allowed
momentum ${\bf k}$. However, in a finitely large space with a
finite scale length, say, $a$, radiation fields with ${\bf k}$
less than $\sim \frac{\pi}{a}$ does not exist in this space.

In order to investigate the projection of spin operators in a
comoving system onto the fixed frame, we first discuss the
expression for the polarization vectors ${\vec\varepsilon}(k,1)$
and ${\vec\varepsilon}(k,2)$ in terms of the 3-D Cartesian
orthogonal unit vectors ${\bf i}$, ${\bf j}$, ${\bf k}$ in the
fixed frame of reference, {\it i.e.},
\begin{equation}
{\vec\varepsilon}(k,1)=e_{i}{\bf i}+e_{j}{\bf j}+e_{k}{\bf k},
\quad {\vec\varepsilon}(k,2)=f_{i}{\bf i}+f_{j}{\bf j}+f_{k}{\bf
k}.       \label{eq26}
\end{equation}
Thus one can arrive at
\begin{eqnarray}
a(k,1)&=&a_{i}(k,1)+a_{j}(k,1)+a_{k}(k,1)        \nonumber \\
&=&i\frac{1}{\sqrt{2(2\pi)^{3}\omega}}\left\{{\int{{\rm d}^{3}{\bf%
x}\exp(i{ k\cdot x})\overline{\partial_{0}}e_{i}A_{i}}+\int{{\rm%
d}^{3}{\bf x}\exp(i{k\cdot
x})\overline{\partial_{0}}e_{j}A_{j}}+\int{{\rm d}^{3}{\bf
x}\exp(i{k\cdot x})\overline{\partial_{0}}e_{k}A_{k}}}\right\},
\label{eq27}
\end{eqnarray}
where $\overline{\partial_{0}}$ is defined to be
$A\overline{\partial_{0}}B=A\partial_{0} B-(\partial_{0} A )B$. In
the same fashion,
\begin{equation}
a^{\dagger}(k,2)=a_{i}^{\dagger}(k,2)+a_{j}^{\dagger}(k,2)+a_{k}^{\dagger}(k,2).
\label{eq28}
\end{equation}
According to Eq.(\ref{eq27}) and (\ref{eq28}), we obtain the
following commuting relations
\begin{equation}
[a_{i}(k,\lambda),a_{j}^{\dagger}(k',\lambda')]=\delta^{3}({\bf
k}-{\bf
k}')\delta_{ij}\varepsilon_{i}(k,\lambda)\varepsilon_{j}(k',\lambda'),
\quad
[a_{i}(k,\lambda),a_{j}(k',\lambda')]=[a_{i}^{\dagger}(k,\lambda),a_{j}^{\dagger}(k',\lambda')]=0.
\end{equation}
Set
\begin{equation}
a_{i}(k,\lambda)=\varepsilon_{i}(k,\lambda)b_{i}(k,\lambda), \quad
a_{i}^{\dagger}(k,\lambda)=\varepsilon_{i}(k,\lambda)b_{i}^{\dagger}(k,\lambda),
\label{eq210}
\end{equation}
and one can readily obtain
\begin{equation}
[b_{i}(k,\lambda),b_{j}^{\dagger}(k',\lambda')]=\delta^{3}({\bf
k}-{\bf k}')\delta_{ij}, \quad
[b_{i}(k,\lambda),b_{j}(k',\lambda')]=[b_{i}^{\dagger}(k,\lambda),b_{j}^{\dagger}(k',\lambda')]=0.
\end{equation}
Note that in (\ref{eq210}) the repeated indices $i$'s does not
imply the summations over them. By the aid of (\ref{eq26}) and
(\ref{eq210}), one can arrive at
\begin{eqnarray}
a^{\dagger}(k,1)a(k,2)-a^{\dagger}(k,2)a(k,1)&=&(e_{i}f_{j}-e_{j}f_{i})[b_{i}^{\dagger}(k)b_{j}(k)-b_{j}^{\dagger}(k)b_{i}(k)]    \nonumber \\
&+&(e_{j}f_{k}-e_{k}f_{j})[b_{j}^{\dagger}(k)b_{k}(k)-b_{k}^{\dagger}(k)b_{j}(k)]+(e_{k}f_{i}-e_{i}f_{k})[b_{k}^{\dagger}(k)b_{i}(k)-b_{i}^{\dagger}(k)b_{k}(k)].
\end{eqnarray}
For convenience, the photon momentum ${\bf k}$ can be expressed in
terms of angle displacements $\lambda$ and $\gamma$ in the
spherical coordinate system, {\it i.e.}, $\frac{{\bf
k}}{k}=(\sin\lambda \cos\gamma, \sin\lambda \sin \gamma,
\cos\lambda)$. So,

\begin{equation}
e_{i}f_{j}-e_{j}f_{i}=\cos\lambda, \quad
e_{j}f_{k}-e_{k}f_{j}=\sin\lambda \cos\gamma,   \quad
e_{k}f_{i}-e_{i}f_{k}=\sin\lambda \sin\gamma.
\end{equation}
The projection of photon spin ${\bf S}$ onto the photon momentum
${\bf k}$ is therefore written
\begin{eqnarray}
I&\equiv&\frac{{\bf k}}{k}\cdot{\bf
S}=-i[a^{\dagger}(k,1)a(k,2)-a^{\dagger}(k,2)a(k,1)]
=-i\{\cos\lambda[b_{i}^{\dagger}(k)b_{j}(k)-b_{j}^{\dagger}(k)b_{i}(k)]    \nonumber \\
&+&\sin\lambda\cos\gamma[b_{j}^{\dagger}(k)b_{k}(k)-b_{k}^{\dagger}(k)b_{j}(k)]+\sin\lambda\sin\gamma[b_{k}^{\dagger}(k)b_{i}(k)-b_{i}^{\dagger}(k)b_{k}(k)]\},
\label{eq214}
\end{eqnarray}
which is referred to as the photon helicity. It follows from
(\ref{eq214}) that the spin operator in the comoving frame
projected onto ${\bf k}$ is just equal to the projection of the
following operator vector
\begin{equation}
{\bf S}_{\rm
fix}=-i(b_{j}^{\dagger}(k)b_{k}(k)-b_{k}^{\dagger}(k)b_{j}(k),\quad
b_{k}^{\dagger}(k)b_{i}(k)-b_{i}^{\dagger}(k)b_{k}(k),\quad
b_{i}^{\dagger}(k)b_{j}(k)-b_{j}^{\dagger}(k)b_{i}(k))
\label{eq215}
\end{equation}
onto ${\bf k}$. It is thus concluded without any fear that the
operator vectors ${\bf S}_{\rm fix}$ can be considered the spin
operator vectors in the fixed frame. Moreover, it is readily
verified that ${\bf S}_{\rm fix}$ agrees with the algebraic
commuting relation of angular momentum operators, {\it i.e.},
\begin{equation}
{\bf S}_{\rm fix}\times{\bf S}_{\rm fix}=i{\bf S}_{\rm fix},
\end{equation}
which confirms that ${\bf S}_{\rm fix}$ in (\ref{eq215}) is truly
an expression for photon spin operator in the fixed frame of
reference. Thus, with the help of
\begin{equation}
[S_{k},\frac{1}{\sqrt{2}}(b_{i}^{\dagger}\pm
ib_{j}^{\dagger})]\equiv[-i(b_{i}^{\dagger}b_{j}-b_{j}^{\dagger}b_{i}),\frac{1}{\sqrt{2}}(b_{i}^{\dagger}\pm
ib_{j}^{\dagger})]=\pm\frac{1}{\sqrt{2}}(b_{i}^{\dagger}\pm
ib_{j}^{\dagger}),
\end{equation}
the eigenstates to the eigenvalue equations $S_{k}|\pm, k>=\pm
|\pm, k>$ of spin operator $S_{k}\equiv
-i(b_{i}^{\dagger}b_{j}-b_{j}^{\dagger}b_{i})$ is obtained as
follows
\begin{equation}
|\pm, k>=\frac{1}{\sqrt{2}}(b_{i}^{\dagger}(k)\pm
ib_{j}^{\dagger}(k))|0>.
\end{equation}

It is worthwhile to point out that the photon helicity,
$\frac{{\bf k}}{k}\cdot{\bf S}$, is a conserved operator, since it
follows from (\ref{eq214}) that the eigenvalue of $\frac{{\bf
k}}{k}\cdot{\bf S}$ does not vary with time $t$ whether it is
observed from comoving or fixed frames. This, therefore, means
that the photon helicity $\frac{{\bf k}}{k}\cdot{\bf S}$ can be
thought of a Lewis-Riesenfeld invariant operator\cite{Lewis}. Here
it is also implied that if the noncoplanarly curved optical fiber
is wound smoothly on a large enough diameter, then the wave vector
of a photon propagating inside the fiber is always along the
tangent of fibre at each point at arbitrary time.
\\ \\

   In accordance with the Lewis-Riesenfeld
theory\cite{Lewis}, the Lewis-Riesenfeld invariant,
$I(t)=\frac{{\bf k}}{k}\cdot{\bf S}$, satisfies the following
Liouville-Von Neumann equation
\begin{equation}
\frac{\partial I(t)}{\partial t}+\frac{1}{i}[I(t),H_{\rm
eff}(t)]=0.                \label{eq219}
\end{equation}
It follows from the above Liouville-Von Neumann equation that one
can obtain the effective Hamiltonian $H_{\rm eff}(t)$ as follows
\begin{equation}
H_{\rm eff}(t)=\frac{{\bf{k}}(t)\times
\dot{\bf{k}}(t)}{k^{2}}\cdot \bf{S} \label{eq220}
\end{equation}
with dot denoting the time derivative. Apparently, insertion of
(\ref{eq220}) into (\ref{eq219}) yields an equation of motion of a
photon
\begin{equation}
{\dot{\bf{k}}}+{\bf{k}}\times ( \frac{{\bf{k}}\times
\dot{\bf{k}}}{k^{2}})=0,         \label{eq100}
\end{equation}
where ${\bf{k}}\times (\frac{\bf{k}\times \dot{\bf{k}}}{k^{2}})$
and $\frac{ \bf{k}\times \dot{\bf{k}}}{k^{2}}$ may be considered
the generalized Lorentz magnetic force (Coriolis force) and the
generalized magnetic field strength\cite{JQ}, respectively. It is
readily verified that Eq.(\ref{eq100}) is an identity. Thus it is
shown that the above mathematical treatment
Eq.(\ref{eq219})-Eq.(\ref{eq100}) for the propagation of photons
inside a curved fibre is self-consistent.

In order to investigate the {\it non-cyclic non-adiabatic}
geometric phases, we should first exactly solve the following
time-dependent Schr\"{o}dinger equation
\begin{equation}
i\frac{\partial \left| \sigma ,{\bf{k}}(t)\right\rangle }{\partial t}=\frac{%
{\bf{k}}(t)\times \dot{\bf{k}}(t)}{k^{2}}\cdot {\bf{S}}\left|
\sigma ,{\bf{k}}(t)\right\rangle,         \label{eq221}
\end{equation}
which governs the time evolution of photon fields propagating
inside the noncoplanarly curved optical fibre\cite{Shen1,Gao},
where $\sigma$ represents the helicity eigenvalue, {\it i.e.}, the
eigenvalue of the invariant $I(t)\equiv\frac{{\bf k}}{k}\cdot{\bf
S}$.

Note that neither the Hamiltonian nor the consequent
time-evolution equation similar to (\ref{eq221}) existed in the
Chiao and Wu's original work\cite{Chiao}. Instead, there was the
following eigenvalue equation
\begin{equation}
\frac{{\bf k}}{k}\cdot{\bf S}\left| \sigma
,{\bf{k}}(t)\right\rangle=\sigma\left| \sigma
,{\bf{k}}(t)\right\rangle.          \label{eq222}
\end{equation}
The connections and differences between (\ref{eq221}) and
(\ref{eq222}) were seen in references\cite{Shen1,Shen2}. By making
use of Berry's cyclic adiabatic geometric phase
formula\cite{Berry2} and Eq.(\ref{eq222}), Chiao and Wu predicted
successfully the existence of photon Berry's phases in the
fibre\cite{Chiao}. In our previous work\cite{Shen1,Shen2}, we
studied Eq.(\ref{eq221}) (which governs the second-quantized
time-dependent spin model) by using the invariant theory and then
treated the {\it non-cyclic non-adiabatic} geometric phases of
photons in the curved fibre. Moreover, by considering the
non-normal-order spin operators (and hence the effective
Hamiltonian), we calculate the quantum-vacuum geometric phases of
photon fields, which may be a new vacuum effects. It is believed
that this geometric phase at quantum-vacuum level is essentially
significant not only experimentally but also theoretically, which
will be illustrated in the Concluding Remarks.

By complicated lengthy calculations, the solution to the
time-dependent Schr\"{o}dinger equation (\ref{eq221}) is obtained
as follows\cite{Shen1}
\begin{equation}
\left| {\bf{k}}(t)\right\rangle=\sum_{\sigma }C_{\sigma }\exp
[\frac{1}{i}\phi _{\sigma }^{\rm (g)}(t)]V(t)\left| \sigma,k
\right\rangle,
\end{equation}
where $\left| \sigma,k \right\rangle\equiv\left|
\sigma,{\bf{k}}(t=0) \right\rangle$ is the initial photon
polarized state, and the time-independent coefficients $C_{\sigma
}=\langle \sigma ,t=0\left| \sigma ,{\bf{k}}(t=0)\right\rangle$
and $V(t)=\exp
[\beta (t)S_{+}-\beta ^{\ast }(t)S_{-}]$ with the time-dependent parameters $\beta (t)=-\frac{\lambda (t)}{2}\exp [-i\gamma (t)],\quad \beta ^{\ast }(t)=-\frac{%
\lambda (t)}{2}\exp [i\gamma (t)]$. The geometric phase of photons
whose initial helicity eigenvalue is $\sigma $ can be expressed by

\begin{equation}
\phi _{\sigma }^{\rm
(g)}(t)=\left\{{\int_{0}^{t}\dot{\gamma}(t^{^{\prime }})[1-\cos
\lambda (t^{^{\prime }})]{\rm d}t^{^{\prime
}}}\right\}\left\langle \sigma,k \right| S_{3}\left| \sigma,k
\right\rangle .            \label{eq224}
\end{equation}
In the adiabatic process where both the precessional frequency
$\dot{\gamma}$ (expressed by $\Omega$) and $\lambda$ (${\bf k }$
deviating from the third axis in the fixed frame by an angle
$\lambda$) can be regarded as constants, the adiabatic geometric
phase ({\it i.e.}, Berry's topological phase) in a cycle
($T=\frac{2\pi}{\Omega}$) in the momentum ${\bf k}$ space is
written
\begin{equation}
\phi _{\sigma }^{\rm (g)}(T)=2\pi(1-\cos \lambda )\left\langle
\sigma,k \right| S_{3}\left| \sigma,k \right\rangle,
                                                        \label{eq225}
\end{equation}
where $2\pi(1-\cos \lambda )$ is equal to a solid angle subtended
at the origin of momentum ${\bf k}$ space. This fact thus means
that the geometric phase (\ref{eq224}) or (\ref{eq225}) carries
information on the global and topological properties of time
evolution of quantum systems. Geometric phases is hence of
physical interest in a wide variety of fields. It should be
emphasized that here the {\it quantum-vacuum geometric phase} may
also be involved in $\left\langle \sigma,k \right| S_{3}\left|
\sigma,k \right\rangle$ if the third component $S_{3}$ of photon
spin operator ${\bf S}$ is of a non-normal-order form, which will
be taken into account in the next section.

Here we omit the derivation of solving Eq.(\ref{eq221}) and only
quote the results. For the detailed derivation procedure the
reader is directed to the reference\cite{Shen1}.
\section{Further discussing quantum-vacuum geometric phases of photons in the curved fibre}
In this section we will further discuss the quantum-fluctuation
contributions to geometric phases of photons moving inside a
sufficiently perfect optical fibre. In the previous section, I
deal only with the normal-order photon spin operators
(\ref{eq22}), which does not involve the vacuum zero-point
electromagnetic fluctuation fields. So, the previous mathematical
treatment cannot predict the existence of {\it quantum-vacuum
geometric phases} of photons in the curved fibre. In order to
treat the so-called {\it quantum-vacuum geometric phases}, we
should study the non-normal-order spin operators, where the
zero-point electromagnetic fields is involved in the effective
Hamiltonian (\ref{eq220}) (and hence in the time-evolution
equation (\ref{eq221})).

    Readers may be referred to the references
\cite{Chiao,Tomita,Kwiat,Ross,Haldane1,Jiao,Ryder,Zhou,Robinson,Haldane2,Berry}
for the early investigations of {\it adiabatic cyclic} geometric
phases (Berry's topological phases) of photons inside a curved
fibre. Historically, the similar work has also been seen in a
remarkable paper published in 1941 in which Vladimirskii had
treated this global topological phase problem in an extension of
an earlier paper published in 1938 by Rytov\cite{Berry}. In all
these researches, authors treated the photon geometric phases in
the coiled optical fibre by making use of the Maxwell's
electrodynamics, differential geometry method as well as Berry's
adiabatic quantum theory. However, by constructing a
second-quantized effective Hamiltonian, we considered the {\it
non-cyclic non-adiabatic} (rather than {\it cyclic adiabatic})
geometric phases of photons inside a noncoplanarly curved
fibre\cite{Shen1,Shen2,Gao} based on the Lewis-Riesenfeld
invariant theory\cite{Lewis} and the invariant-related unitary
transformation formulation\cite{Gao1}. In the paper\cite{Shen1},
we exactly solved some time-dependent quantum models by using the
invariant theory and then studied in detail the time-evolution
operator of photon wavefunction in the curved fibre (here the
wavefunction time-evolution operator is an exact solution to the
time-dependent Schr\"{o}dinger equation, rather than that
associated with the chronological product). In the
paper\cite{Shen2}, we investigated both non-adiabatic geometric
phases and helicity reversals (as well as some related topics) of
photons propagating inside the coiled fibre and briefly considered
the potential applications of photon helicity inversion to the
communication and information science. In this published
work\cite{Shen2}, we also suggested the quantum-vacuum geometric
phases of photons in the fibre\footnote{As far as we are
concerned, the photon propagation problem can be ascribed to a
time-dependent second-quantized spin model, where the effective
(phenomenological) Hamiltonian ({\it e.g.}, the expression
(\ref{eq220})) is of a second-quantized form. Whereas in the
previous
researches\cite{Chiao,Tomita,Kwiat,Ross,Haldane1,Jiao,Ryder,Zhou,Robinson,Haldane2,Berry},
this problem was treated often by using classical Maxwell's
Equations and first-quantized Schr\"{o}dinger equation (and
Berry's adiabatic geometric phase formula as well\cite{Berry2}).
Although these investigations can be said to be somewhat
outstandingly successful in both predicting and studying adiabatic
geometric phases of photons in the fibre, here I still want to
emphasize two points: for one thing, only by using the
second-quantization formulation can we investigate the photon
geometric phases at quantum level; for another, only when we
consider the non-normal-product second-quantized Hamiltonian can
it enable us to predict the existence of geometric phases at
quantum-vacuum level. Tomita and Chiao may also agree to the above
first point. They held the arguments\cite {Haldane2} that although
the geometric phases in the curved fibre can also be obtained by
means of classical Maxwell's electrodynamics, they preferred to
think of this phenomenon as originating at the quantum level, but
surviving the correspondence-principle limit into the classical
level. However, this point is not the main subject in the present
paper, which will be further discussed elsewhere. Here, instead,
we concentrate only on the second point, {\it i.e.}, the geometric
phases at quantum-vacuum level resulting from the zero-point
radiation fields of vacuum, which has not been investigated in
previous researches.}, which is a physically interesting concept
and might perhaps focus attention of researchers in various
fields, where the quantum systems are of second quantization.

Substituting the Fourier expansion series (\ref{eeqA}) of {\bf
A({\bf x},t)} into the expression (\ref{eq21}) for photon spin
operator, one can obtain the non-normal-order photon ${\bf S}$
\cite{Shen1}, {\it i.e.},
\begin{equation}
{\bf S}=\frac{i}{2}\frac{{\bf
 k}}{k}[a(k,1)a^{\dagger}(k,2)-a^{\dagger}(k,1)a(k,2)-a(k,2)a^{\dagger}(k,1)+a^{\dagger}(k,2)a(k,1)].        \label{eq31}
\end{equation}
In what follows we define the creation and annihilation operators,
$a_{R}^{\dagger}(k)$, $a_{L}^{\dagger}(k)$, $a_{R}(k)$,
$a_{L}(k)$, of right- and left- handed circularly polarized light
\cite{Bjorken}
\begin{eqnarray}
a_{R}^{\dagger}(k)&=&\frac{1}{\sqrt{2}}[a^{\dagger}(k,1)+ia^{\dagger}(k,2)],
\quad
a_{R}(k)=\frac{1}{\sqrt{2}}[a(k,1)-ia(k,2)],           \nonumber \\
a_{L}^{\dagger}(k)&=&\frac{1}{\sqrt{2}}[a^{\dagger}(k,1)-ia^{\dagger}(k,2)],
\quad a_{L}(k)=\frac{1}{\sqrt{2}}[a(k,1)+ia(k,2)].
\label{eq32}
\end{eqnarray}
It follows that
\begin{eqnarray}
a^{\dagger}(k,1)&=&\frac{1}{\sqrt{2}}[a_{R}^{\dagger}(k)+a_{L}^{\dagger}(k)],
\quad
a(k,1)=\frac{1}{\sqrt{2}}[a_{R}(k)+a_{L}(k)],      \nonumber \\
a^{\dagger}(k,2)&=&\frac{1}{\sqrt{2}i}[a_{R}^{\dagger}(k)-a_{L}^{\dagger}(k)],
\quad a(k,2)=-\frac{1}{\sqrt{2}i}[a_{R}(k)-a_{L}(k)].
\end{eqnarray}
So, the monomode-photon spin operator (\ref{eq31}) can be
rewritten
\begin{equation}
{\bf S}=\frac{1}{2}\frac{{\bf
 k}}{k}\left\{{[a_{R}(k)a_{R}^{\dagger}(k)+a_{R}^{\dagger}(k)a_{R}(k)]-[a_{L}(k)a_{L}^{\dagger}(k)+a_{L}^{\dagger}(k)a_{L}(k)]}\right\}.
\end{equation}
Thus, according to the definition of photon helicity $I(t)
\equiv\frac{{\bf k}}{k}\cdot {\bf S} $, $I(t)$ is given by
\begin{equation}
I(t)=\frac{1}{2}\left\{{[a_{R}(k)a_{R}^{\dagger}(k)+a_{R}^{\dagger}(k)a_{R}(k)]-[a_{L}(k)a_{L}^{\dagger}(k)+a_{L}^{\dagger}(k)a_{L}(k)]}\right\}.
\end{equation}
Hence we can construct the left- and right- handed photon states
as follows
\begin{equation}
|\sigma=-1,k\rangle=a_{L}^{\dagger}(k)|0\rangle, \quad
 |\sigma=+1,k\rangle=a_{R}^{\dagger}(k)|0\rangle,              \label{eq36}
\end{equation}
where the two helicity eigenvalue equations are satisfied
\begin{equation}
I(t)|\sigma=-1,k\rangle=-|\sigma=-1,k\rangle, \quad
I(t)|\sigma=+1,k\rangle=+|\sigma=+1,k\rangle.        \label{eq37}
\end{equation}
Note that the above discussion is performed in the comoving frame
of reference. In what follows, we consider the spin operator in a
fixed frame. In accordance with the invariant-related unitary
transformation formulation\cite{Gao,Shen1}, the {\it
time-dependent} invariant $I(t)$ can be transformed into a {\it
time-independent} operator
\begin{equation}
I_{\rm V}\equiv V^{†}(t)I(t)V(t)=S_{3},
\end{equation}
where $V(t)=\exp
[\beta (t)S_{+}-\beta ^{\ast }(t)S_{-}]$ with $\beta (t)=-\frac{\lambda (t)}{2}\exp [-i\gamma (t)],\quad \beta ^{\ast }(t)=-\frac{%
\lambda (t)}{2}\exp [i\gamma (t)]$. Here $\lambda(t)$ and
$\gamma(t)$ are angle displacements in the spherical coordinate
system and are so defined $\frac{{\bf k(t)}}{k}=(\sin\lambda (t)
\cos\gamma (t), \sin\lambda (t) \sin \gamma (t), \cos\lambda (t))$
as mentioned in Sec. II. Since $V(t=0)=1$, we obtain $\lambda
(t=0)=0$ and hence $k_{3}=k$ and $k_{1}=k_{2}=0$. Thus in the
Schr\"{o}dinger picture, in which the (initial) photon momentum
$k_{3}=k$ and $k_{1}=k_{2}=0$, the third component of photons spin
operator is therefore of the form
\begin{equation}
 S_{3}=\frac{1}{2}\left\{{[a_{R}(k)a_{R}^{\dagger}(k)+a_{R}^{\dagger}(k)a_{R}(k)]-[a_{L}(k)a_{L}^{\dagger}(k)+a_{L}^{\dagger}(k)a_{L}(k)]}\right\}.                         \label{eq39}
\end{equation}
It can be found from (\ref{eq220}) and (\ref{eq39}) that the
time-dependent zero-point energy exists in the effective
Hamiltonian, which will result in the quantum-vacuum geometric
phases. These aspects are illustrated in the application of
(\ref{eq39}) to the LRH geometric phase formulae which follow. It
should be noted that here $a_{R}(k)$, $a_{R}^{\dagger}(k)$,
$a_{L}$ and $a_{L}^{\dagger}(k)$ are regarded as the
time-independent operators\footnote{Since in Sec. II we have
discussed the photon spin operators projected onto the fixed frame
of reference, it follows from (\ref{eq31}) that the {\it
time-independent} ({\it i.e.}, in the fixed frame or
Schr\"{o}dinger picture) third component of spin $S_{3}$ is
written as
$S_{3}=\frac{i}{2}[b_{1}(k)b_{2}^{\dagger}(k)-b_{1}^{\dagger}(k)b_{2}(k)-b_{2}(k)b_{1}^{\dagger}(k)+b_{2}^{\dagger}(k)b_{1}(k)]$
(non-normal-order). So, the {\it time-independent} creation and
annihilation operators of right- and left- handed circularly
polarized light are respectively defined to be
$a_{R}^{\dagger}(k)=\frac{1}{\sqrt{2}}[b_{1}^{\dagger}(k)+ib_{2}^{\dagger}(k)]$,
 $a_{R}(k)=\frac{1}{\sqrt{2}}[b_{1}(k)-ib_{2}(k)]$ and
$a_{L}^{\dagger}(k)=\frac{1}{\sqrt{2}}[b_{1}^{\dagger}(k)-ib_{2}^{\dagger}(k)]$,
 $a_{L}(k)=\frac{1}{\sqrt{2}}[b_{1}(k)+ib_{2}(k)]$. The monomode-photon states corresponding
 to helicity eigenvalues $\sigma=\pm 1$ are therefore constructed in terms of $b_{1}^{\dagger}(k)$,
 $b_{2}^{\dagger}(k)$, $b_{1}(k)$, $b_{2}(k)$ and vacuum state $|0\rangle$ as follows: $|\sigma=\pm
1,k\rangle=\frac{1}{\sqrt{2}}[b_{1}^{\dagger}\pm
ib_{2}^{\dagger}]|0\rangle$. Actually, according to the analysis
in Sec. II, the definition of left- and right- handed polarized
photon states in (\ref{eq36}) should be replaced with the
above-defined $|\sigma=\pm
1,k\rangle=\frac{1}{\sqrt{2}}[b_{1}^{\dagger}\pm
ib_{2}^{\dagger}]|0\rangle$. But, for convenience, here we do not
make a difference between them, since under the initial condition
$k_{3}=k$, $k_{1}=k_{2}=0$, these two definitions of
monomode-photon polarized states corresponding
 to helicity eigenvalues $\sigma=\pm 1$ are consistent with each other.}.

The monomode multi-photon states of left- and right- handed (LRH)
circularly polarized light (at $t=0$) can be defined
\begin{equation}
|\sigma=-1,k,
n_{L}\rangle=\frac{[a_{L}^{\dagger}(k)]^{n}}{\sqrt{n!}}|0_{L}\rangle,
\quad
|\sigma=+1,k,
n_{R}\rangle=\frac{[a_{R}^{\dagger}(k)]^{n}}{\sqrt{n!}}|0_{R}\rangle
\label{eq310}
\end{equation}
with $n_{L}$ and $n_{R}$ being the LRH polarized photon occupation
numbers, respectively. In the following we calculate the geometric
phases of multi-photon states
\begin{equation}
|\sigma=+1,k, n_{R}; \sigma=-1,k, n_{L}\rangle\equiv|\sigma=+1,k,
n_{R}\rangle\otimes|\sigma=-1,k, n_{L}\rangle \label{eq3100}
\end{equation}
in the fibre. Substitution of (\ref{eq3100}) into (\ref{eq224})
yields
\begin{equation}
\phi^{\rm (g)}(t)=\left\{{\int_{0}^{t}\dot{\gamma}(t^{^{\prime
}})[1-\cos \lambda (t^{^{\prime }})]{\rm d}t^{^{\prime
}}}\right\}\left\langle \sigma=+1,k, n_{R}; \sigma=-1,k, n_{L}
\right| S_{3}|\sigma=+1,k, n_{R}; \sigma=-1,k, n_{L}\rangle .
\label{eq311}
\end{equation}
and the final result is given
\begin{equation}
 \phi^{\rm (g)}(t)=(n_{R}-n_{L})\left\{{\int_{0}^{t}\dot{\gamma}(t^{^{\prime
}})[1-\cos \lambda (t^{^{\prime }})]{\rm d}t^{^{\prime
}}}\right\},                                 \label{eq312}
\end{equation}
which is independent of $k$ but dependent instead on the geometric
nature of the pathway (expressed in terms of $\lambda$ and
$\gamma$) along which the light wave propagates. This fact
indicates that geometric phases possesses the topological and
global properties of time evolution of quantum systems. It is
worth notifying that the phases (\ref{eq312}) associated with the
photonic occupation numbers $n_{R}$ and $n_{L}$ are quantal in
character\cite{Gao}. Gao has shown why $\phi^{\rm
(g)}(t)=(n_{R}-n_{L})\{{\int_{0}^{t}\dot{\gamma}(t^{^{\prime
}})[1-\cos \lambda (t^{^{\prime }})]{\rm d}t^{^{\prime }}}\}$ is
referred to as the quantal geometric phases\cite{Gao} by taking
into account the uncertainty relation between the operators
$\frac{1}{2}[a_{R(L)}^{\dagger}(k)+a_{R(L)}(k)]$ and
$\frac{i}{2}[a_{R(L)}^{\dagger}(k)-a_{R(L)}(k)]$. Although the
phases $\phi^{\rm (g)}(t)$ in (\ref{eq312}) are quantal geometric
phases of photons, they do not belong to the geometric phases at
quantum-vacuum level which arise, however, from the zero-point
electromagnetic energy of vacuum quantum fluctuation.

It should be noted that the cyclic adiabatic cases of
(\ref{eq312}) (or $n_{R}\{{\int_{0}^{t}\dot{\gamma}(t^{^{\prime
}})[1-\cos \lambda (t^{^{\prime }})]{\rm d}t^{^{\prime }}}\}$ and
$-n_{L}\{{\int_{0}^{t}\dot{\gamma}(t^{^{\prime }})[1-\cos \lambda
(t^{^{\prime }})]{\rm d}t^{^{\prime }}}\}$ alone) have been
measured experimentally by Tomita and Chiao {\it et
al.}\cite{Tomita,Kwiat,Ross,Haldane1}.

According to the expression (\ref{eq39}) for ${\bf S_{3}}$, both
the geometric phases of left- and right- handed circularly
polarized photon states, {\it i.e.}, $|\sigma=-1,k, n_{L}\rangle$
and $|\sigma=+1,k, n_{R}\rangle$, are respectively of the form
\begin{equation}
 \phi_{L}^{\rm (g)}(t)=-(n_{L}+\frac{1}{2})\left\{{\int_{0}^{t}\dot{\gamma}(t^{^{\prime
}})[1-\cos \lambda (t^{^{\prime }})]{\rm d}t^{^{\prime
}}}\right\},                \quad
 \phi_{R}^{\rm (g)}(t)=+(n_{R}+\frac{1}{2})\left\{{\int_{0}^{t}\dot{\gamma}(t^{^{\prime
}})[1-\cos \lambda (t^{^{\prime }})]{\rm d}t^{^{\prime
}}}\right\}.                 \label{eq313}
\end{equation}

It follows that the time-dependent zero-point energy possesses
physical meanings and therefore contributes to geometric phases of
photon fields. Thus the non-cyclic non-adiabatic geometric phases
of left- and right- handed polarized states at quantum-vacuum
level are given
\begin{equation}
 \phi_{\sigma=\pm 1}^{\rm (vacuum)}(t)=\pm\frac{1}{2}\left\{{\int_{0}^{t}\dot{\gamma}(t^{^{\prime
}})[1-\cos \lambda (t^{^{\prime }})]{\rm d}t^{^{\prime
}}}\right\}.                       \label{eq314}
\end{equation}
Note that the geometric phases expressed in (\ref{eq314}) possess
quantal belongings and, moreover, has no classical counterpart (or
correspondence), namely, it cannot survive the
correspondence-principle limit into the classical level.

Compared (\ref{eq312}) with (\ref{eq314}), it is easily seen that
geometric phases (\ref{eq312}) arises from the real photons
(having close relation to the occupation numbers of left- and
right- handed polarized photons), while the quantum-vacuum phases
(\ref{eq314}) results from the vacuum electromagnetic fluctuation
(virtual photons). This quantum-vacuum nature in the latter case
makes $\phi_{\sigma=\pm 1}^{\rm (vacuum)}(t)$ more physically
interesting and therefore the experimental observation for
$\phi_{\sigma=\pm 1}^{\rm (vacuum)}(t)$ deserves consideration.

However, it should be pointed out that, unfortunately, even at the
quantum level, this observable quantum-vacuum geometric phases
$\phi_{\sigma=\pm 1}^{\rm (vacuum)}(t)$ is absent in the fibre
experiment, since it follows from (\ref{eq313}) and (\ref{eq314})
that the signs of quantal geometric phases of left- and
right-handed circularly polarized photons are just opposite to one
another, and so that their quantum-vacuum geometric phases are
counteracted by each other. Hence the observed geometric phases
are only those expressed by (\ref{eq312}), of which whose
adiabatic case has been measured in the optical fibre experiment
performed by Tomita and Chiao {\it et
al.}\cite{Tomita,Kwiat,Ross,Haldane1}.
 \\ \\

It is of important significance to investigate the structures and
properties of vacuum in field theory, which has attracted
considerable attention of many investigators, as evidenced by a
considerable amount of literature. Vacuum possesses many effects,
such as Casimir's effect\cite{Casimir} reflecting the zero-point
energy of vacuum, vacuum polarization leading to Lamb's shift
(hyperfine structure of Hydrogen atomic spectra)\cite{Bethe},
atomic spontaneous radiation due to the interaction between the
excited atom and the zero-point electromagnetic field, and the
anomalous magnetic moment of electron as well. Vacuum, as the
ground states of quantum fields, has universal properties of
symmetry. Increasing evidences such as the facts that Nambu and
Goldstone discovered the vacuum spontaneously symmetrical
breaking\cite{Goldstone} of field theory in 1960's, and Polyakov
{\it et al.} found the topological structures of
vacuum\cite{Belavin} in 1970's, have demonstrated that vacuum
possesses abundant properties and deserve detailed investigations.
Field theory ever encounters problems such as divergent zero-point
energies of quantized electromagnetic fields and infinite electric
charge density arising from the presence of electrons of negative
energies. These problems can be solved by taking the normal order
for the field operators, which is consistent with Lorentz
covariance. The background charges and zero-point energies is thus
removed, which makes the vacuum expectation values of both charge
density and Hamiltonian vanish (namely, the infinite constant is
harmless and easily removed by measuring all energies relative to
the vacuum state). In such systems of field theory, of which whose
Hamiltonian is {\it time-independent}, the same amount of
zero-point energy is eliminated at different time, which is
equivalent to re-defining the background energies. Hence, in these
cases the normal-product procedure is both reliable and practical.
However, whether the normal product is valid or not should still
be taken into consideration for systems with {\it time-dependent}
Hamiltonians. Since the time-dependent system is no longer
Lorentz-invariant, evidences for the validity of normal product
are insufficient so far. In other words, if the normal product is
applied to the time-dependent systems of quantum field theory, and
thus the vacuum background is so re-defined by removing different
zero-point energies at different time, then some observable vacuum
effects ({\it e.g.}, Berry's phase) may be cancelled theoretically
and the validity of this formalism therefore deserves incredulity.
In view of the above remarks, it is emphasized that investigations
of vacuum states of time-dependent systems may become particularly
important. So, we think the test of the above quantum-vacuum
geometric phases is now of imperative necessity.

Additionally, why do we say the geometric phase (\ref{eq314}) has
physical meanings? The reasons are as follows: for the first, it
is a time-dependent phase. It is well known that only the
time-dependent phase factor in quantum mechanical wavefunction has
physical meanings. In other words, the time-independent or
constant phase factor lacks physical meanings; for the second,
this phase (\ref{eq314}) is caused by the zero-point fluctuation
fields of vacuum. So, it receives our much attention.
 \\ \\

Unfortunately, since the zero-point LRH polarized electromagnetic
fields are present often accompanied by each other in the
isotropic-media curved fibre, as a matter of fact the
quantum-vacuum geometric phases (\ref{eq314}) cannot be detected
easily in the previous fiber
experiments\cite{Tomita,Kwiat,Ross,Haldane1}\footnote{It follows
from (\ref{eq39}) that the vanishing total LRH quantum-vacuum
geometric phases $\phi_{L}^{\rm (vacuum)}(t)+\phi_{R}^{\rm
(vacuum)}(t)$ is related close to the fact that the total angular
momentum of zero-point electromagnetic fields is vanishing.
Actually, the two problems have quite a lot in common. From the
physical point of view, the total spin of LRH radiation fields of
vacuum vanishes, so does the total LRH quantum-vacuum geometric
phases.}. Although their total effects is vanishing ({\it i.e.},
$\phi_{L}^{\rm (vacuum)}(t)+\phi_{R}^{\rm (vacuum)}(t)=0$), the
quantum-vacuum geometric phase of left- or right- handed
circularly polarized photon state still truly exists (but just be
cancelled by each other). The problem now we encounter is: how can
we extract experimentally one of the non-vanishing quantum-vacuum
geometric phases from the vanishing $\phi_{L}^{\rm
(vacuum)}(t)+\phi_{R}^{\rm (vacuum)}(t)$ ?

We think this may prove to the physicists that the resolution of
this problem is truly imperative necessary. The relevant
discussions and remarks are presented in the Appendix to this
paper.
\section{Concluding remarks}
In this paper, we deal with the projection problem of spin
operator of photon fields in the comoving coordinate systems onto
the fixed frame of reference, and discuss further the
quantum-vacuum geometric phases in the optical fibre. It is shown
that the non-normal order of photon spin operators yields
observable effects ({\it e.g.}, quantum-vacuum geometric phases)
arising from zero-point vacuum energies. These two vacuum
geometric phases are exactly equal but different only by a minus
sign. This, therefore, implies that the total LRH quantum-vacuum
geometric phases become exactly zero. So, the only retained
geometric phases are those in (\ref{eq312}), which result from the
occupation numbers of LRH photons in polarized states.
 \\ \\

The physical significance of quantum-vacuum geometric phases of
photon fields in the curved fiber may be given as follows:

(i) the photon geometric phases at quantum-vacuum level originates
from the zero-point electromagnetic fluctuations. Since geometric
phases indicates topological and global properties of quantum
systems in time-evolution processes, the quantum-vacuum geometric
phases of photons in the helically wound fibre may contain the
information on the topological and global properties of time
evolution of vacuum fluctuation fields and (we hope it) might
capture attention in a wide variety of fields;

(ii) quantum-vacuum geometric phases may be considered a new
physically interesting vacuum effects. It is known that Casimir's
effect is realized by changing the mode structure of zero-point
electromagnetic fields between the two parallel conducting metal
plates\cite{Casimir} ({\it i.e.}, by modifying the magnitude
distribution of electromagnetic wave vector {\bf k}). However, the
quantum-vacuum geometric phases of photons is produced by altering
the direction of wave vector {\bf k} in the curved fibre ({\it
i.e.}, photon fields itself alters its wave vector {\bf k} when
travelling along the curved fibre);

(iii) the problem of whether the normal-product procedure is valid
or not for time-dependent quantum systems still remains unclear so
far. We believe that the consideration of this problem may enable
us to investigate the time-dependent quantum field theory. If, for
example, when we study the quantum field theory in the curved
space-time, where one should often deal with the particle creation
problem in the time-dependent gravitational
backgrounds\cite{Fulling,Ford}, we think the vacuum effects
associated with evolution process of time-dependent quantized
field systems may necessarily also be taken into consideration.
So, from the point of view of us, it is important to test
experimentally the quantum geometric phases so as to answer the
question mentioned above. If we cannot find the existence of this
geometric phases at quantum-vacuum level ({\it i.e.}, this vacuum
effect does not exist), then it is believed that the
normal-product procedure in second quantization is still valid and
correct for time-dependent quantum systems. But, if the
quantum-vacuum geometric phases is present experimentally, then we
argue that the normal-product procedure in second quantization may
be invalid and should be discussed further when treating
time-dependent quantum systems. We think it might be a leading
problem in both quantization formulation and vacuum physics. This
problem is under consideration and will be published elsewhere. We
are longing to perform a relevant experiment to confirm our above
interpretations.

However, it is most unfortunate that the two quantum-vacuum
geometric phases of LRH polarized photons for each ${\bf k}$ are
eliminated by each other, and furthermore no one among them
appears to want to be readily extracted either.
\\ \\

\textbf{Acknowledgements} I thank X.C. Gao for his helpful
proposals. This project is supported in part by the National
Natural Science Foundation of China under the project No.
$90101024$.
\\ \\

\begin{appendix}
\textbf{APPENDIX}

Can We Extract the Quantum-vacuum Geometric Phases from {\bf
NOTHING}?
\\ \\

It follows from (\ref{eq314}) that the total vacuum phases ({\it
i.e.}, $\phi_{L}^{\rm (vacuum)}(t)+\phi_{R}^{\rm (vacuum)}(t)$)
vanishes. Here our aim is to resolve the problem of how to extract
experimentally the quantum-vacuum geometric phases $\phi_{L}^{\rm
(vacuum)}(t)$ or $\phi_{R}^{\rm (vacuum)}(t)$ from the vanishing
total vacuum geometric phases.

Our brief history of investigating photon geometric phases in the
fibre is as follows: in April 2000, Gao and I began to consider
the {\bf non-cyclic non-adiabatic geometric phases} of photon
fields in the curved fibre based on a {\bf second-quantized spin
model} (see spin model, for example, in
references\cite{Bouchiat,Datta,Mizrahi,GaoXC,Zhu}). In May 2000,
Gao first proposed the concept of {\bf quantum-vacuum geometric
phases}. The existence problem of quantum-vacuum geometric phases
is strongly relevant to whether the second quantization in spin
model is adopted or not. Since it is in connection with properties
of quantum electromagnetic vacuum and, moreover, this geometric
phase is related close to the topological and global features of
time evolution of vacuum-fluctuation fields, we think this concept
is of essential significance and therefore deserves detailed
investigations. From then on, these problems gained our attention
and we tried to investigate this topological phases at
quantum-vacuum level.

Since quantum-vacuum geometric phases has an important connection
with vacuum energies, these experimental realizations may be
relevant to the validity problem of normal-product procedure in
the time-dependent quantum field theory (TDQFT), {\it i.e.}, we
also aim to re-examine the normal-product procedure in some
extensions. If the {\it quantum-vacuum geometric phases} is proved
present experimentally, then it is reasonably believed that it is
not suitable for us to remove vacuum fluctuation energies and
infinite charge density just by using the old formulation, {\it
e.g.}, re-defining the vacuum background energies and electric
charges by utilizing the normal-product procedure, since in this
re-definition, some potential physically interesting vacuum
effects may also be removed theoretically. We think only for the
{\it time-independent} quantum field systems can we use safely the
normal-product procedure without any fear of introducing any new
problems other than those which quantum field theory had
encountered before\cite{Bjorken2}. However, for the {\it
time-dependent} quantum field systems, ({\it e.g.}, photon fields
propagating inside a helically curved fibre and quantum fields in
an expanding universe), the physically interesting vacuum effects
will unfortunately be deducted by the second-quantization
normal-order formulation. So, the normal-order technique may
therefore not be applicable to the {\it time-dependent} quantum
fields. To the best of our knowledge, in the literature, this
normal order problem in the time-dependent quantum field theory
gets less attention and interests than it deserves. To test our
above theoretical viewpoints, we hope the quantum-vacuum geometric
phases of photons in the curved fibre would be investigated
experimentally in the near future.

Unfortunately, the left-handed polarized light due to vacuum
fluctuation is often accompanied by the zero-point right-handed
polarized light and their total quantum-vacuum geometric phases is
therefore vanishing. So, it is not easy for physicists to
investigate experimentally the quantum-vacuum geometric phases.
This, therefore, means that our above theoretical remarks as to
whether the normal-product procedure is valid or not for the
time-dependent quantum field theory (TDQFT) cannot be examined
experimentally. During the last three years, I tried my best but
unfortunately failed to suggest an excellent idea of experimental
realization of quantum-vacuum geometric phases. We conclude that
it seems not quite satisfactory to test the quantum-vacuum
geometric phases by using the optical fibre that is made of
isotropic media, inhomogeneous media ({\it e.g.}, photonic
crystals\footnote{Photonic crystals are artificial materials
patterned with a periodicity in dielectric constant, which can
create a range of forbidden frequencies called a photonic band
gap. Such dielectric structure of crystals offers the possibility
of molding the flow of light (including the zero-point
electromagnetic fields of vacuum). It is believed that, in the
similar fashion, this effect ({\it i.e.}, modifying the mode
structures of vacuum electromagnetic fields) may also take place
in gyrotropic media. This point will be applied to the discussion
which follows.}), left-handed media (a kind of artificial
composite metamaterial with negative refractive index), uniaxial
(biaxial) crystals or chiral materials. Is it truly extremely
difficult to realize such a goal? It is found finally that perhaps
in the fibre composed of some anisotropic media (such as
gyrotropic materials, including also gyroelectric and gyromagnetic
media, where both electric permittivity and magnetic permeability
are respectively the tensors) the quantum-vacuum geometric phases
may be achieved test experimentally. In these gyrotropic media,
only one of the LRH polarized lights can be propagated without
being absorbed by media. This result holds also for the zero-point
electromagnetic fields. It is well known that people can
manipulate vacuum so as to alter the zero-point mode structures of
vacuum, which has been illustrated in photonic crystals and
Casimir's effect (additionally, the space between two parallel
mirrors, cavity in cavity QED, {\it etc.}). If, for example, in
some certain gyrotropic media one of the LRH polarized lights,
say, the left-handed polarized light, dissipates due to the medium
absorption and only the right-handed light is allowed to be
propagated (in the meanwhile the mode structure of vacuum in these
anisotropic media also alters correspondingly), then the
quantum-vacuum geometric phase of right-handed polarized light can
be easily tested in the fibre fabricated from these gyrotropic
media.

To close this Appendix, we consider briefly the electromagnetic
wave equations in a homogeneous gyroelectric medium, where the
electric permittivity tensor, $(\epsilon)_{ik}$, and magnetic
permeability (scalar) are given

\begin{equation}
(\epsilon)_{ik}=\left(\begin{array}{cccc}
\epsilon_{1}  & i\epsilon_{2} & 0 \\
-i\epsilon_{2} &   \epsilon_{1} & 0  \\
 0 &  0 &  \epsilon_{3}
 \end{array}
 \right),                 \quad            \mu=\mu.              \eqnum{A1}
 \label{A1}
\end{equation}
Assuming that the direction of the electromagnetic wave vector
${\bf k}$ is parallel to the third component of the Cartesian
coordinate system, with the help of Maxwell's Equations, we obtain
the following two planar wave equations
\begin{eqnarray}
 \nabla^{2}[\frac{1}{\sqrt{2}}({\bf E}_{1}+i{\bf E}_{2})]-\epsilon_{0}\mu_{0}\mu(\epsilon_{1}+\epsilon_{2})\frac{\partial^{2}}{\partial t^{2}}[\frac{1}{\sqrt{2}}({\bf E}_{1}+i{\bf E}_{2})]=0,  \nonumber  \\
 \nabla^{2}[\frac{1}{\sqrt{2}}({\bf E}_{1}-i{\bf E}_{2})]-\epsilon_{0}\mu_{0}\mu(\epsilon_{1}-\epsilon_{2})\frac{\partial^{2}}{\partial t^{2}}[\frac{1}{\sqrt{2}}({\bf E}_{1}-i{\bf
 E}_{2})]=0,       \eqnum{A2}
\end{eqnarray}
of which whose optical refractive indices squared take the form
\begin{equation}
 n_{\pm}^{2}=\mu(\epsilon_{1}\pm \epsilon_{2}),                 \eqnum{A3}
 \label{A3}
\end{equation}
where the sign $\pm$ corresponds to the two directions of
polarization vectors of light wave.

For simplicity, without loss of generality, it is assumed that the
two mutually perpendicular real unit polarization vectors
${\vec\varepsilon}(k,1)$ and ${\vec\varepsilon}(k,2)$ can be taken
to be as follows: $\varepsilon_{1}(k,1)=\varepsilon_{2}(k,2)=1$,
$\varepsilon_{1}(k,2)=\varepsilon_{2}(k,1)=0$ and
$\varepsilon_{3}(k,1)=\varepsilon_{3}(k,2)=0$. Thus by the aid of
(\ref{eq a}), (\ref{eq32}) and the formula ${\bf
E}=-\frac{\partial {\bf A}}{\partial t}$ for the electric field
strength, in the second-quantization framework one can arrive at
\begin{eqnarray}
\frac{1}{\sqrt{2}}({\bf E}_{1}+i{\bf E}_{2})=i\int{\rm d}^{3}{\bf
k}\sqrt{\frac{\omega}{2(2\pi)^{3}}}[a_{L}(k)\exp(-ik\cdot
x)-a_{R}^{\dagger}(k)\exp(ik\cdot x)],   \nonumber  \\
\frac{1}{\sqrt{2}}({\bf E}_{1}-i{\bf E}_{2})=i\int{\rm d}^{3}{\bf
k}\sqrt{\frac{\omega}{2(2\pi)^{3}}}[a_{R}(k)\exp(-ik\cdot
x)-a_{L}^{\dagger}(k)\exp(ik\cdot x)]. \eqnum{A4} \label{A4}
\end{eqnarray}
We now discuss the wave propagation in a gyroelectric-medium
optical fibre. As an illustrative example, we think only of the
condition under which
$|\epsilon_{2}|>|\epsilon_{1}|=-\epsilon_{1}, \mu>0$. If, for
instance, $\epsilon_{2}$ is positive, then
$\frac{1}{\sqrt{2}}({\bf E}_{1}+i{\bf E}_{2})$ can be propagated
while $\frac{1}{\sqrt{2}}({\bf E}_{1}-i{\bf E}_{2})$ cannot be
propagated (because of the negative $n_{-}^{2}$ and the consequent
imaginary propagation constant $k_{-}$, which is expressed by
$(n_{-})\frac{\omega}{c}$) in the fibre; conversely,
$\frac{1}{\sqrt{2}}({\bf E}_{1}-i{\bf E}_{2})$ can be propagated
while $\frac{1}{\sqrt{2}}({\bf E}_{1}+i{\bf E}_{2})$ is inhibited
from being propagated (due to the imaginary propagation constant
$k_{+}$) if $\epsilon_{2}$ is negative. Thus in the former case
the phase $\phi_{R}^{\rm (vacuum)}(t)$ of right-handed polarized
light and in the latter case, instead the phase $\phi_{L}^{\rm
(vacuum)}(t)$ of left-handed polarized light, may respectively be
detected in the gyroelectric-medium fibre experiments. It is
accepted by us that in resolving this problem ({\it i.e.},
experimental realizations of quantum-vacuum geometric phases) it
is necessarily of much more practical importance to think of
gyrotropic media than of any other inhomogeneous and chiral
materials. We hope the simple analysis ({\it i.e.}, from
(\ref{A1}) to (\ref{A4})) would prove useful to physicists for
investigating quantal geometric phases of electromagnetic fields
in the fibre (of course, indeed, Eq.(\ref{A3}) {\it etc.} have
provided us with some insights into these problems!).

We are now in a position to further investigate some new
properties of this potential vacuum effect anticipated for such
time-dependent zero-point fields of vacuum (or a quantum system in
time-dependent environments such as a time-dependent homogeneous
electric field\cite{Fu1}, time-dependent gravitational
backgrounds\cite{Fu2} and expanding space-time), particularly if
we could perform some experiments to test the presented viewpoints
in Sec. III and the Appendix. We hope this might offer some new
insights into the physics of the problems mentioned above (in
particular, the validity of normal product in time-dependent
quantum field theories should be further clarified).
 \\ \\

 {\bf NOTE}: In order to
illustrate the content in the present paper, an outline is given
as follows:
\\ \\

 \setlength{\unitlength}{1cm}
\begin{picture}(30,5)
\put(0,4){\framebox(10.7,0.5){Chiao-Wu's model (photons moving
inside a helically curved fibre)}} \put(10.9,4){\vector(1,0){3.8}}
\put(11.1,4.1){Berry's phase formula}
\put(0,3.2){\framebox(9.1,0.5){cyclic adiabatic geometric phase
$\phi _{\sigma }^{\rm (g)}(T)=2\pi\sigma(1-\cos \lambda )$}}
\put(9.3,3.2){\vector(1,0){5.5}} \put(9.5,3.3){Lewis-Riesenfeld
invariant theory} \put(0,2.4){\framebox(12.3,0.5){non-cyclic
non-adiabatic geometric phase $\phi _{\sigma }^{\rm
(g)}(t)=\sigma\{{\int_{0}^{t}\dot{\gamma}(t^{^{\prime }})[1-\cos
\lambda (t^{^{\prime }})]{\rm d}t^{^{\prime }}}\}$}}
\put(12.5,2.4){\vector(1,0){4.8}} \put(12.7,2.5){second-quantized
spin model}

\put(0,1.6){\framebox(11.2,0.5){quantal geometric phases
$\phi^{\rm
(g)}(t)=(n_{R}-n_{L})\{{\int_{0}^{t}\dot{\gamma}(t^{^{\prime
}})[1-\cos \lambda (t^{^{\prime }})]{\rm d}t^{^{\prime }}}\}$}}
\put(11.4,1.6){\vector(1,0){4.1}} \put(11.6,1.7){under non-normal
order}

\put(0,0.8){\framebox(10.8,0.6){quantum-vacuum phases
$\phi_{\sigma=\pm 1}^{\rm
(vacuum)}(t)=\pm\frac{1}{2}\{{\int_{0}^{t}\dot{\gamma}(t^{^{\prime
}})[1-\cos \lambda (t^{^{\prime }})]{\rm d}t^{^{\prime }}}\}$}}
\put(11,0.8){\vector(1,0){7.1}} \put(11.1,0.9){validity problem of
normal order in TDQFT}

\put(0,0){\framebox(4.6,0.5){experimental test is required}}
\put(4.8,0){\vector(1,0){7.2}} \put(4.9,0.15){unfortunately,
$\phi_{L}^{\rm (vacuum)}(t)+\phi_{R}^{\rm (vacuum)}(t)=0$
}\put(12.2,0){\framebox(5.3,0.6){by using gyrotropic-medium
fibre}}
\end{picture}

\end{appendix}

\end{document}